\documentclass[journal]{IEEEtran}
\usepackage{graphicx}
\usepackage{multirow}
\usepackage{amsmath}
\usepackage{subfigure}
\usepackage{cite}
\usepackage{colortbl}
\definecolor{Gray}{gray}{0.85}
\usepackage{xcolor}
\usepackage{multirow}

\ifCLASSINFOpdf
\else
\fi
\begin{document}
%
\title{CMOS Circuit Implementation of Spiking Neural Network for Pattern Recognition Using On-chip Unsupervised STDP Learning }
%
%
%

\author{Sahibia~Kaur~Vohra,~\IEEEmembership{Student~Member,~IEEE,}
        Sherin~A~Thomas,~\IEEEmembership{Student~Member,~IEEE,}
        Mahendra~Sakare,~\IEEEmembership{Member,~IEEE,}
        and~Devarshi~Mrinal~Das,~\IEEEmembership{Senior~Member,~IEEE}
\thanks{Sahibia Kaur Vohra, Sherin A Thomas, Mahendra Sakare and Devarshi Mrinal Das are with the Department of Electrical Engineering, Indian Institute of Technology, Ropar, Rupnagar 140001, India (e-mail: sahibia.19eez0002@iitrpr.ac.in, sherin.19eez0001@iitrpr.ac.in, mahendra@iitrpr.ac.in, devarshi.das@iitrpr.ac.in).}}
\maketitle

\begin{abstract}
Computation on a large volume of data at high speed and low power requires energy-efficient computing architectures.
Spiking neural network (SNN) with bio-inspired spike-timing-dependent plasticity learning (STDP) is a promising solution for energy-efficient neuromorphic systems than conventional artificial neural network (ANN). Previous works on SNN with STDP learning primarily uses memristive devices which are difficult to fabricate. Some reported works on SNN makes use of memristor macro models, which are software-based and cannot give complete insight into circuit implementation challenges. This article presents for the first time, a full circuit-level implementation of the SNN system featuring on-chip unsupervised STDP learning in standard CMOS technology. It does not involve the use of FPGAs, CPUs or GPUs for training the neural network. We demonstrated the complete circuit-level design, implementation and simulation of SNN with on-chip training and inference for pattern classification using 180 nm CMOS technology. A comprehensive comparison of the proposed SNN circuit with the previous related work is also presented. To demonstrate the versatility of the CMOS synapse circuit for application scenarios requiring rate-based learning, we have tuned the pair-based STDP circuit to obtain Bienenstock-Cooper-Munro (BCM) characteristics and applied it to heart rate classification.
\end{abstract}


%
\IEEEpeerreviewmaketitle

\section{Introduction}
%
%
%
%
\IEEEPARstart{S}{piking} neural network (SNN) is the third generation artificial neural network that provides a promising solution for replacing area and power-hungry hardware for neuromorphic computing. The brain's superior energy efficiency for decision-making cognitive tasks made scientists to focus their efforts on building non-Von Neumann computer systems that imitate the biological brain. Neurons process information as asynchronous event-driven spikes and retain memories as synaptic strengths of their connection in the brain. In this regard, a spiking neural network (SNN) is more bio-plausible than other neural networks that can pave a new way for future intelligent devices for low-power computing applications.

\begin{figure}[!t]
	\centering
	\includegraphics[scale=0.43,trim={0.0 0.68cm 0 0.3cm},clip]{"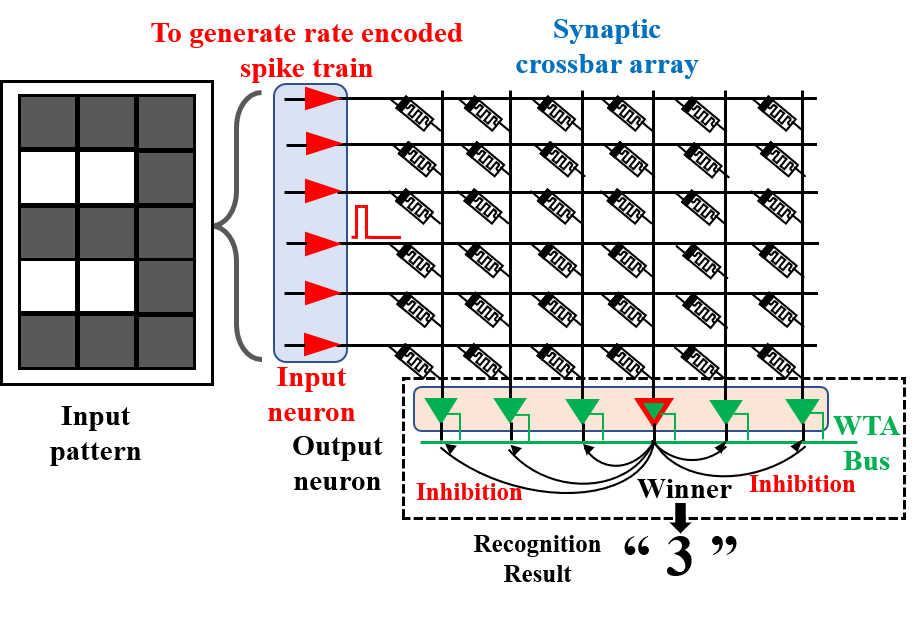"}
	\caption{The simplified architecture of SNN for pattern recognition comprising input layer LIF neurons, memristive crossbar and output layer LIF neurons with WTA mechanism.}
	\label{SNN_arch}
\end{figure}

Spiking neuromorphic computing architecture as shown in Fig.~\ref{SNN_arch}, processes information in the form of spikes. Therefore, input sensory analog signal should be converted to spikes which can be performed by various neuron models \cite{c3}. Out of all the neuron models, the LIF neuron model gives a good balance between accuracy and ease of hardware implementation, and also it resembles much of
biological neurons \cite{c3,b9}. The other important block of 
the SNN is the synapse circuit that stores the synaptic weight and defines the strength of the connection between the neurons. Memristor proves to be the most suitable candidate for emulating a synapse as it provides the tunable and non-volatile storage of synaptic weights \cite{b1,b33}. Memristive neuromorphic systems outperform Von-Neumann systems in power efficiency and learning capabilities \cite{c1}.

The learning mechanism used for updating the synaptic weights is a crucial aspect of the neural network. The training algorithm such as backpropagation used in literature \cite{31,32} is widely established but needs extensive hardware resources and hard to be fully implemented in analog circuits. Many works have shown the training of memristive neural network using the ex-situ method \cite{34,35} where an external circuit or computational platform is required for weight calculation. In contrast to these learning methods, a bio-plausible learning method spike-timing-dependent plasticity (STDP) can be used. It has been proved that STDP can be used to train the SNN in-situ with unsupervised learning without compromising parallelism \cite{b3,N2,12,tnano2,tnano1}.
Inference in SNN by taking pre-trained memristor array consumes significantly less power; however, training the synaptic weights of the SNN before inference efficiently remains challenging \cite{1,2}. The approach of training the non-SNN first and then converting it to an SNN has several limitations in terms of accuracy loss and large inference latency \cite{1,3}. The more efficient way is to do both training and inference in SNN hardware. Some works have shown the training by mapping the values of the obtained weight using learning algorithms in software to the neural crossbar \cite{N1}. Some make use of GPUs, FPGAs or microcontrollers, which increases the complexity and power consumption \cite{N1,46}. As a result, doing the training on the implemented hardware for SNN (known as in-situ training) is the more power-efficient way.

The realisation of neuromorphic circuits with memristive synapse have shown wider applications for low power and energy and area-efficient computing neuromorphic system-on-a-chip (NeuSoc) implementation. However, the varying threshold of the memristive devices, stochastic switching and variable resistance states put challenges on the memristive devices \cite{f11}. In addition, unlike CMOS-based circuits, such memristive devices require additional fabrication steps \cite{9}, which makes them challenging to integrate with standard CMOS circuit components in commercial foundries. The memristive devices are yet not available in any standard CMOS technology PDK. As a result, while memristive device technology matures, CMOS-based memristive synapse circuits can be investigated for real-time hardware implementation of neuromorphic circuits. Many works \cite{35,a3,a4,a1} show the SNN architecture for pattern recognition applications using the SPICE model of the memristor, which cannot be used for real hardware implementation. These challenges and limitations motivated us to design an SNN system with CMOS memristor emulators to get full insight into the circuit implementation challenges for memristive SNN based neuromorphic computing (NMC) system. In work \cite{v3,v2} we have used the CMOS memristor emulator given in \cite{b4} for implementing the full CMOS memristive neural network.

The main contributions of this work are:
\begin{enumerate}
	\item A complete CMOS based SNN system for pattern recognition is designed. To the best of our knowledge, the complete CMOS based circuit design and implementation of an analog SNN system with memristor emulator circuit as a synapse has not been reported yet.
	\item The proposed SNN system shows the unsupervised on-chip STDP learning and inference with transistor-level circuit implementation, which is not reported yet in literature. The functionality of the proposed system is demonstrated by a pattern recognition task.
	\item To demonstrate the versatility of the CMOS memristive synapse circuit, we have tuned the CMOS based STDP circuit by exploiting the rate-based Bienenstock-Cooper-Munro (BCM) learning characteristics for heart rate classification.
\end{enumerate}

The rest of the paper is organised as follows. Section II gives the detail of the LIF neuron circuit for converting the input pixel of an image into a spike train. Section III explains the STDP learning obtained using CMOS memristive STDP circuit. Section IV explains the proposed CMOS based SNN system. Section V discusses the simulation, robustness and comparison with the other works. Section VI contains the prospect for the used STDP circuit for heart rate classification application. The conclusion is dawn in section VII.        

\section{LIF Neuron Circuit}
\begin{figure}[!t]
	\centering
	\includegraphics[scale=0.4]{"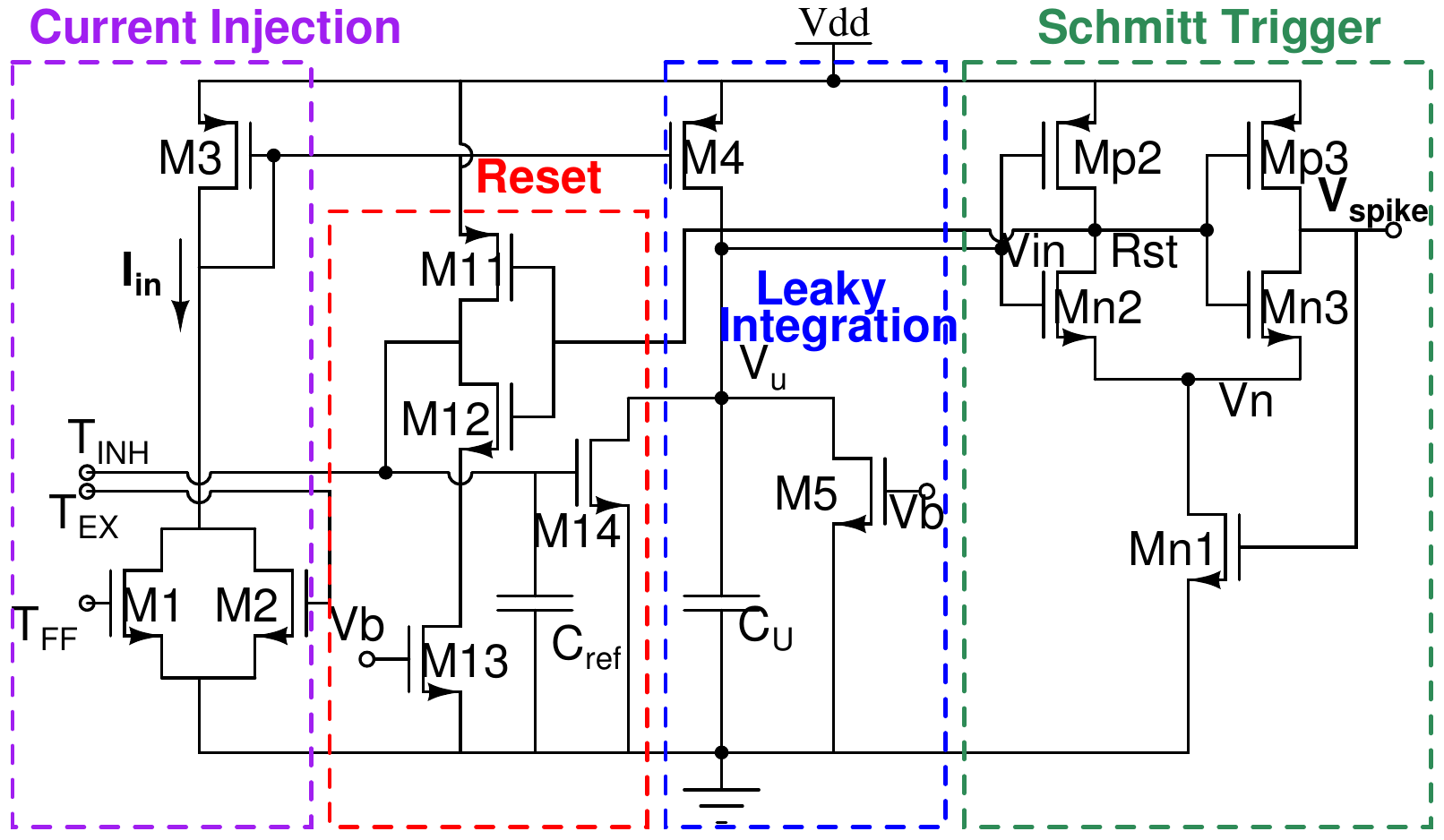"}
	\caption{CMOS circuit of LIF neuron. It comprises various sections for current injection, leaky integration, schmitt trigger for firing and reset [2].}
	\label{LIFschmitt}
\end{figure}

In our SNN system, the low-power, low-complexity neuron circuit proposed in \cite{a3} is employed as a presynaptic and postsynaptic neuron. It contains a lateral inhibition interface, which is used for implementing winner-take-all (WTA) mechanism (refer Fig.~\ref{SNN_arch}) in SNN. In Fig.~\ref{LIFschmitt} input current $I_{in}$, (controlled by voltage given at input terminals $T_{EX}$ and $T_{FF}$) is injected into the current integration section via the current mirror, which charges the capacitor $C_u$.
When membrane voltage $V_u$ reaches the neuron's threshold, the spike is generated with the help of the Schmitt trigger, and simultaneously \textit{Rst} will be low, which turns on M11 and resets the voltage $V_u$ through $C_{ref}$ and M14. The circuit can also get reset through external pin $T_{INH}$. The switching voltage ($V_{SV}$) of the Schmitt trigger \cite{c5} is the neuron's firing threshold, which can be calculated as
\begin{equation} \label{e3}
 I_{Dn2}=\beta_{n2}(V_{SV}-V_n-V_{TH})^2
 \end{equation}
 \begin{equation} \label{e5}
 I_{Dp2}=\beta_{p2}(V_{DD}-V_{SV}-|V_{tp}|)^2
 \end{equation}
 where $I_{Dn,pi}$ is the drain current and $\beta_{n,pi}$ represents the transconductance of transistor $M_{n,pi}$. Now, equating \eqref{e3} and \eqref{e5}, we get
 \begin{equation} \label{e6}
 V_{SW}=V_n + \frac{V_{DD}+V_T(1-R)-V_n}{R+1}
 \end{equation}
 \begin{equation} \label{e9}
 I_{Dn1}=\beta_{n1}(V_n-V_{TH})^2
 \end{equation}
  Equating \eqref{e3} and \eqref{e9}, we get
  \begin{equation} \label{e10}
 V_n=\frac{V_{HL}}{R_n+1}+V_{TH}\frac{R_n-1}{R_n+1}
 \end{equation}
After substituting \eqref{e10} in \eqref{e6}, V$_{SV}$ is given as
 \begin{equation} \label{e11}
 V_{SV}=V_{DD}\frac{R_n+1}{R_n(R+1)+1}+V_{TH}\frac{R_n(2R-1)-1}{R_n(R-1)+1}
 \end{equation}
 where $R=\sqrt{\frac{\beta_{n2}}{\beta_{p2}}}; R_n=\sqrt{\frac{\beta_{n1}}{\beta_{n2}}}$.
Whenever the neuron's membrane voltage ($V_u$) reaches the switching voltage $V_{SV}$, neuron outputs a spike. The width of the spike can be controlled by the switching voltage of the Schmitt trigger given by \eqref{e11}.

\begin{figure}[!t]
	\centering
	\includegraphics[scale=0.35,trim={1cm 1.25cm 0 1.0cm},clip]{"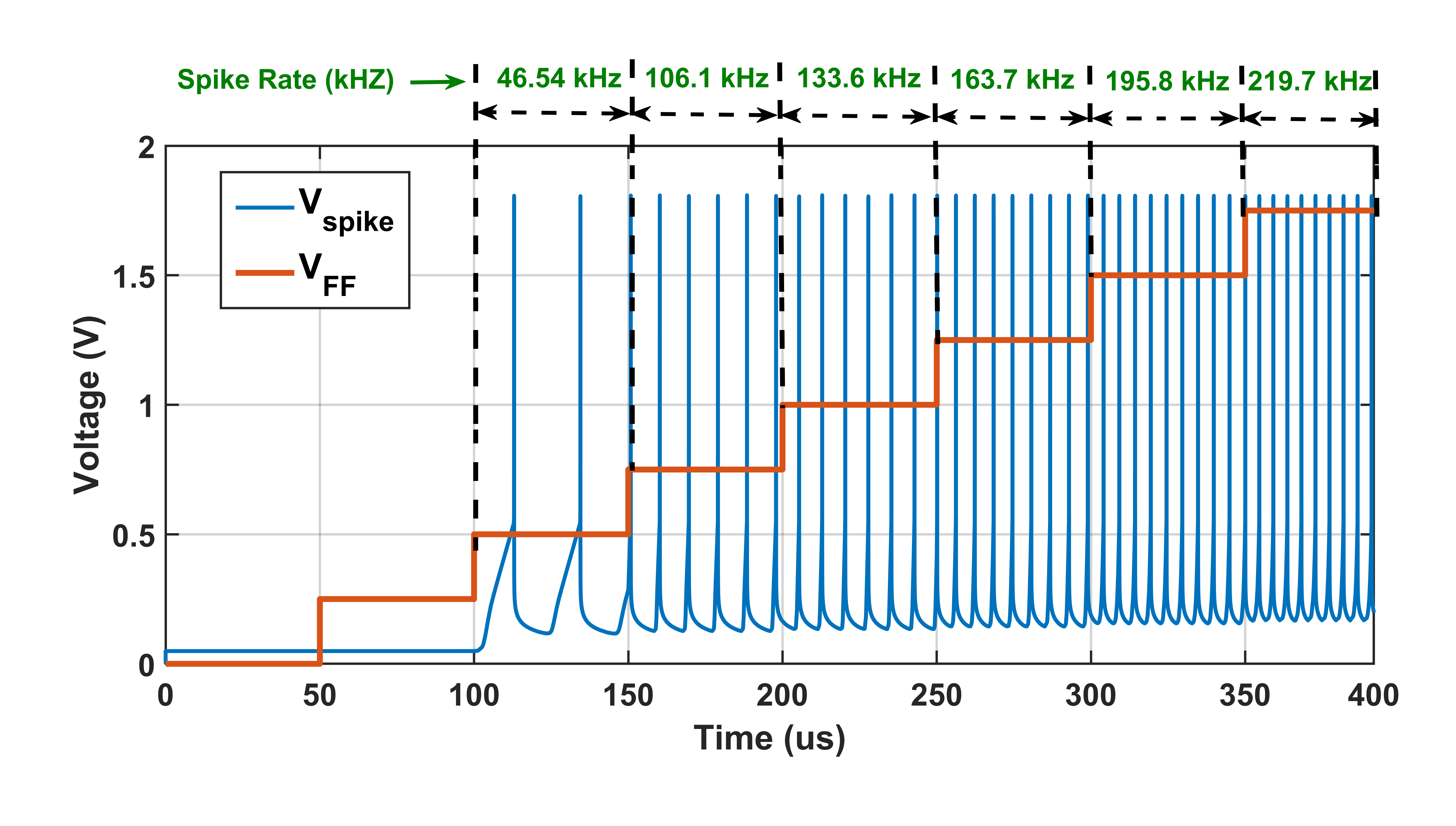"}
	\caption{Response of spiking LIF neuron to different input step voltages. Rate of spiking is increasing with each step of input voltage $V_{FF}$.}
	\label{LIFfreq}
\end{figure}
Further, we have adapted this circuit to meet the desired requirements of our proposed CMOS based SNN system. The simulation results of the LIF circuit implemented in 180 nm CMOS technology is shown in Fig.~\ref{LIFfreq}. It can be observed that the neuron will not generate spikes for input voltage $V_{FF} < 500 mV$  (given at $T_{FF}$) and spiking rate is proportional to the input voltage for  $V_{FF} > 500 mV$. This property is used to convert the input signal into rate encoded spike train.
 \begin{figure}[!t]
	\centering
	\includegraphics[scale=0.35]{"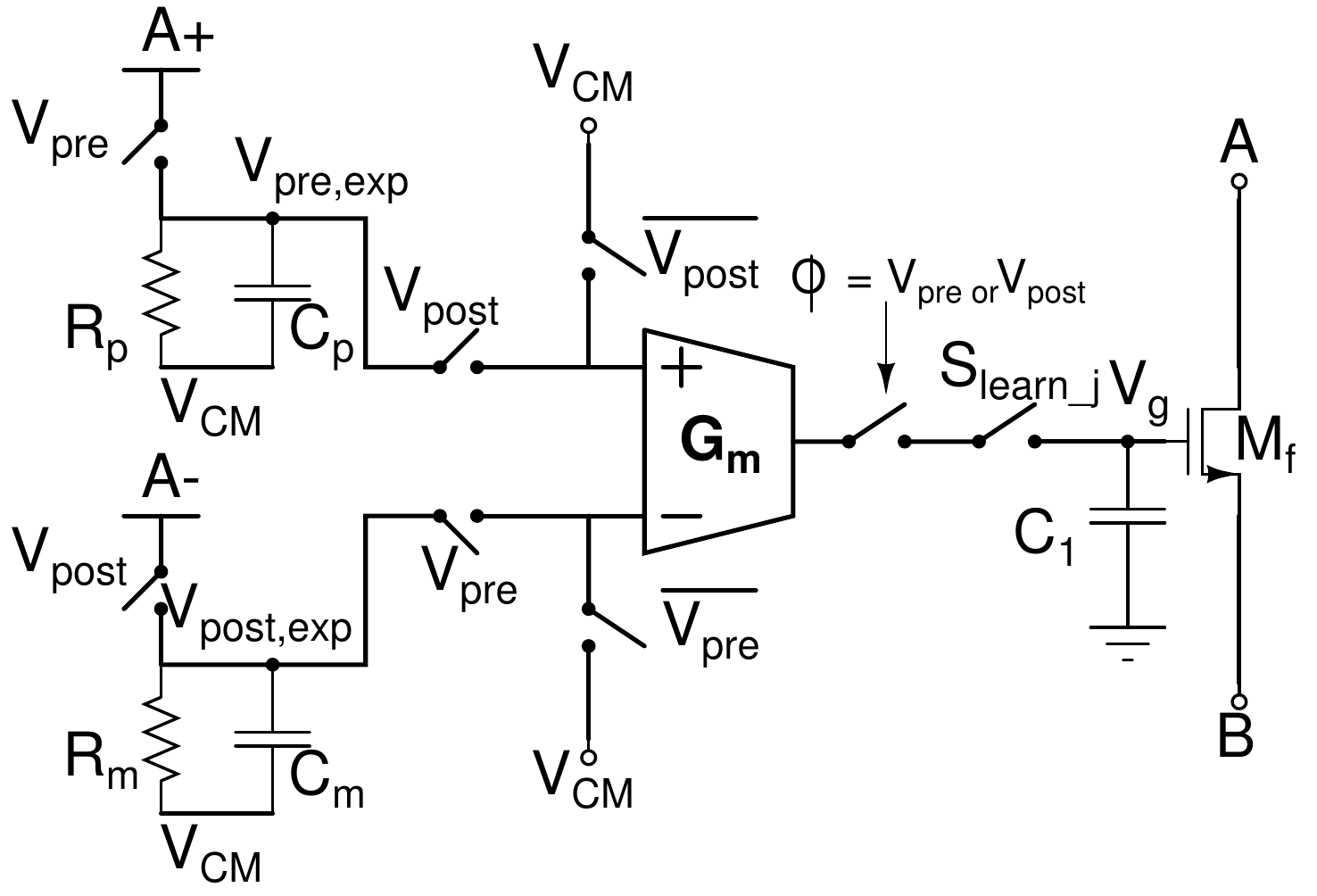"}
	\caption{CMOS circuit of memristive STDP learning synapse.}
	\label{STDP}
\end{figure}
 \begin{table}[!ht]
	\renewcommand{\thefootnote}{\fnsymbol{footnote}}
	\caption{Circuit components and parameter values of STDP synapse}
	\label{table}
	\centering
	\begin{tabular}{|p{1.2cm}|p{1.7cm}||p{1.7cm}|p{1.7cm}|}
		\hline
		\textbf{Name} & \textbf{Values} & \textbf{Name}& \textbf{Values}\\
		\hline
		$R_{p,m}$ & 1M $\Omega$ & HRS & 1.6 M$\Omega$\\
		\hline
	    $C_{p,m}, C_1$ & 1p F & LRS &114 K$\Omega$\\
		\hline
		$A_+,A_-$ & 1.8 V & $G_m$ &18 $\mu$A/V\\
		\hline
		$V_{CM}$ & 900 mV & ${W/L}_{M_f}$ &${0.42/10}\mu$\\
		\hline
	\end{tabular}	
\end{table}
\section{Memristive Synapse Circuit Featuring STDP Learning Mechanism}
\begin{figure}[!t]
	\centering
	\includegraphics[scale=0.31,trim={0.4cm 0.2cm 0 0},clip]{"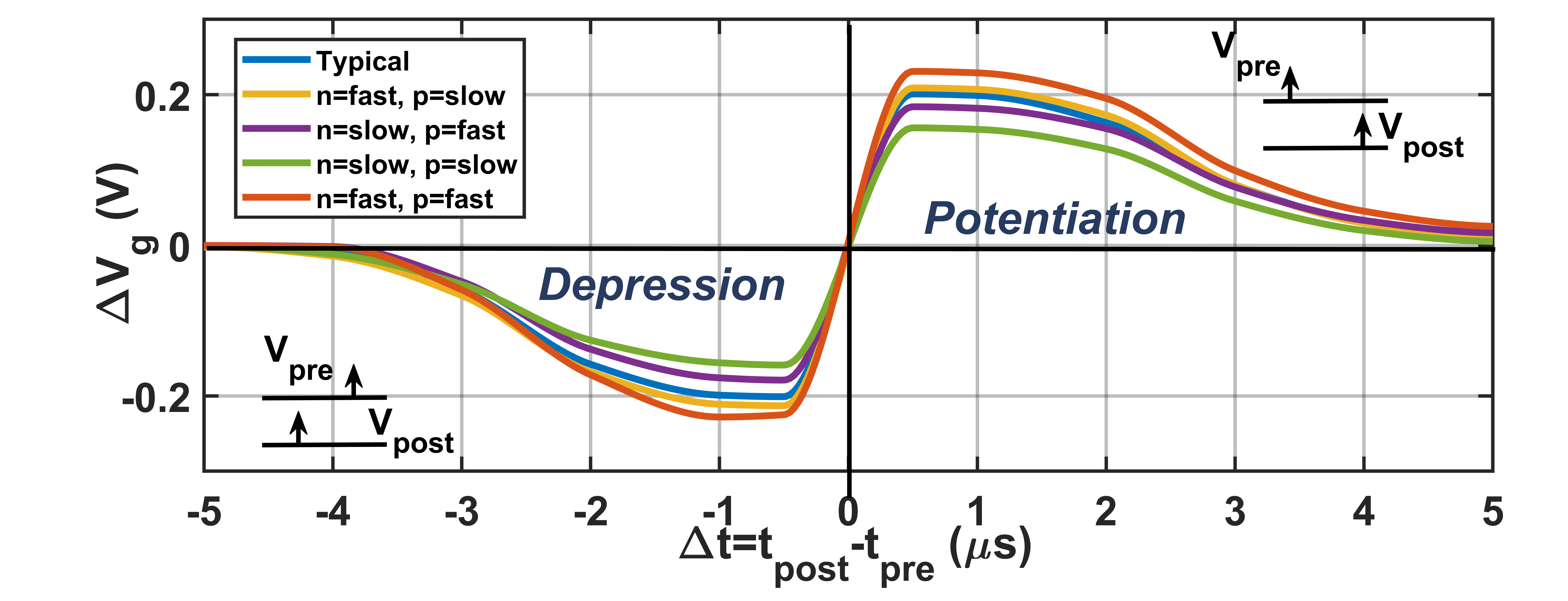"}
	\caption{STDP characterstic curve obtained from the synapse circuit for all process corners with parameters listed in Table I.}
	\label{STDPcurve}
\end{figure}

\begin{figure}[!t]
	\centering
	\includegraphics[scale=0.180,trim={0.1cm 0.3cm 0 1.4cm},clip]{"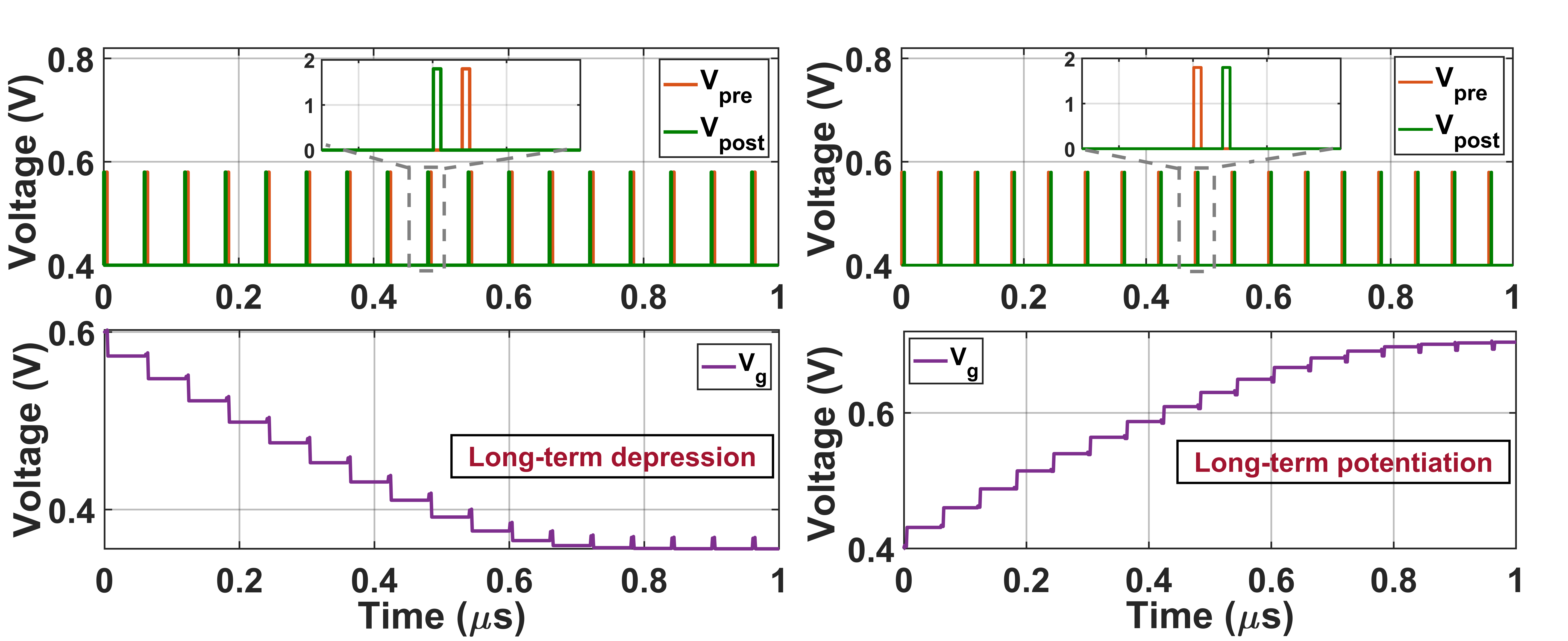"}
	\caption{Transient simulation of a single STDP synapse for Long-term depression (LTD) and long-term potentiation (LTP)}
	\label{LTPLTD}
\end{figure}

STDP learning mechanism observed in biological synapses has the potential to train the SNN in an unsupervised manner \cite{b3}. The phenomenon of STDP learning depends on the spike timing of presynaptic and postsynaptic neuron. It strengthens (weakens) the synaptic weight between two neurons if the presynaptic neuron fires earlier (later) than the postsynaptic neuron. Fig.~\ref{STDP} shows the compact memristive STDP learning circuit given in \cite{a6} which is used as a synapse for our proposed CMOS SNN system. The $R_P$, $C_P$ and $R_m$, $C_m$ shown in Fig.~\ref{STDP} implements the two exponential decay circuits (EDCs) for pre spike and post spike respectively. The post spike (pre spike) samples the $V_{pre,exp}$ $(V_{post,exp})$ trace and update the voltage at the positive (negative) terminal of the operational transconductance amplifier (OTA). The exponential traces are translated to the current through transconductance $G_m$, which will charge the capacitor $C_1$ when the switches (controlled by $\phi, S_{learn_j}$) are closed. The voltage $V_G$ across $C_1$ is the state of the synapse and models the synaptic weight ($W$) by controlling the conductance (\textit{G}) of transistor $M_f$ as given by \eqref{eq10}, \eqref{eq11}
\begin{equation} \label{eq10}
G=K_n\frac{W}{L}(V_g-V_{CM}-V_{THn})
\end{equation}
\begin{equation} \label{eq11}
W=G(V_g)
\end{equation}  
The synapse circuit is implemented in 180 nm CMOS technology with parameters given in Table I. The values of $R_{p,m}$, $C_{p,m}$ have to be selected according to the spiking frequency of neurons so that it can show the synaptic plasticity for the frequency range of the LIF neurons. The simulated result of the synapse circuit shown in Fig.~\ref{STDPcurve} verifies that the STDP learning curve is retained in all process corners. All postsynaptic spikes preceding (succeeding) presynaptic spikes with the delay $\Delta t<t_{max_{pot}}$ ($|\Delta t|<|t_{max_{dep}}|$) result in long-term potentiation (LTP) (Long-term depression (LTD)) as shown in Fig.~\ref{LTPLTD}. $t_{max_{pot}}$ and $|t_{max_{dep}}|$ are the maximum timing difference between presynaptic and postsynaptic spikes after which net change in weight ($\Delta W$) is 0 for potentiation and depression respectively. LTP and LTD lead to the low resistance state (LRS) and high resistance state (HRS) of the STDP memristive synapse, respectively. 
\section{Proposed CMOS based SNN System}
\begin{figure*}[!t]
	\centering
	\includegraphics[scale=0.28]{"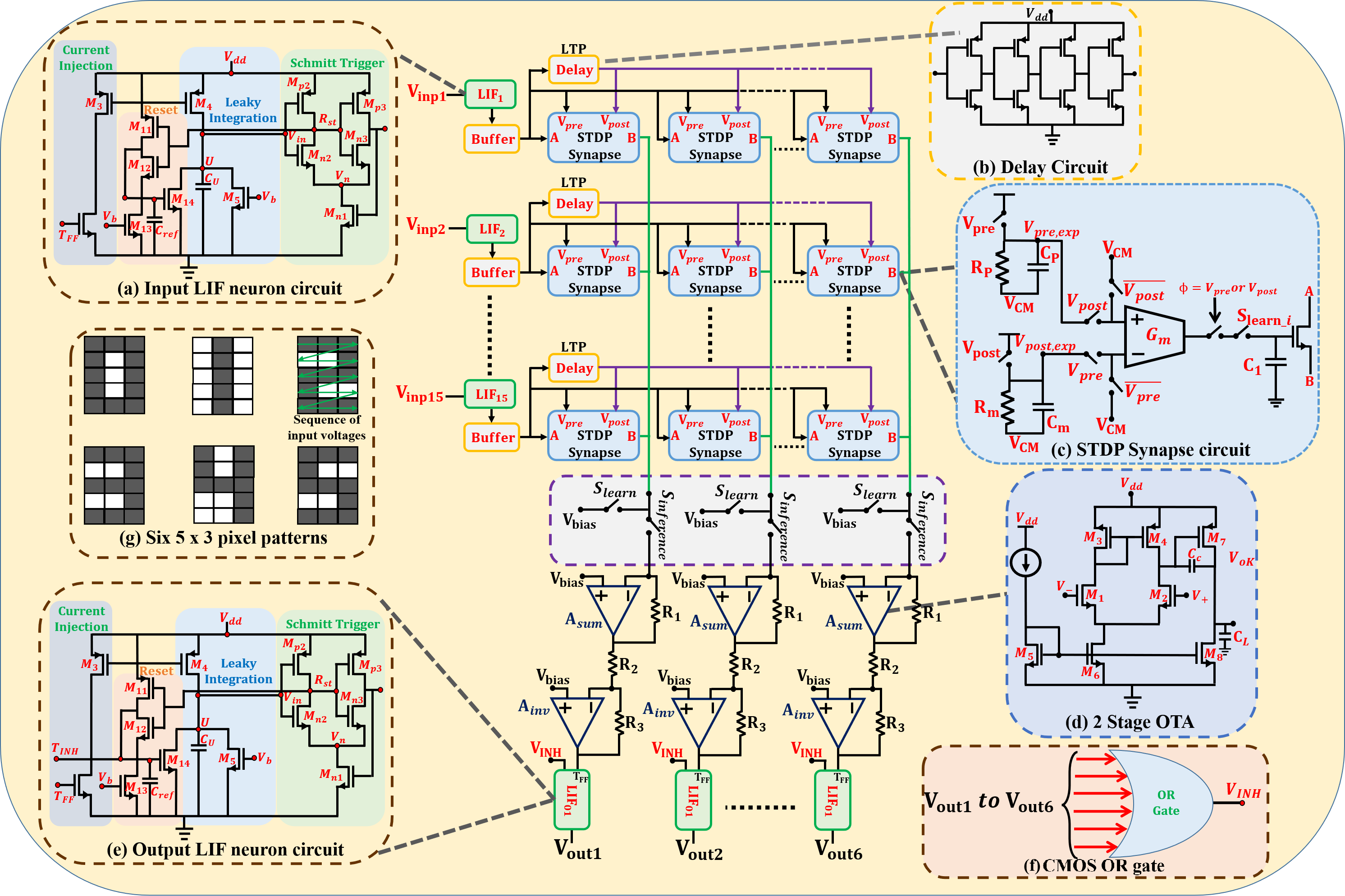"}
	\caption{The proposed CMOS based SNN system architecture combining 15 input LIF neurons, 15$*$6 memristive crossbar array and 6 output LIF neurons. (a) Input LIF neuron circuit. (b) Delay circuit designed for the delay of 1 $\mu$s. (c) CMOS memristive STDP synapse circuit (d) 2-stage OTA used for summing the synaptic currents. (e) Output LIF neuron circuit with inhibitory interface. (f) OR gate for sending the inhibitory signal to output neurons. (g) six 5 X 3 pixels patterns for recognition.}
	\label{SNNsystem}
\end{figure*}

The proposed transistor-level circuit design of unsupervised SNN with in-situ STDP learning is shown in Fig.~\ref{SNNsystem}. The design can be divided into two parts: the circuit design for updating weight (or training phase), and the circuit design for the inference phase (or recognition phase).
\subsection{Circuit Design of Training Phase}
\begin{figure}[!t]
	\centering
	\includegraphics[scale=0.116]{"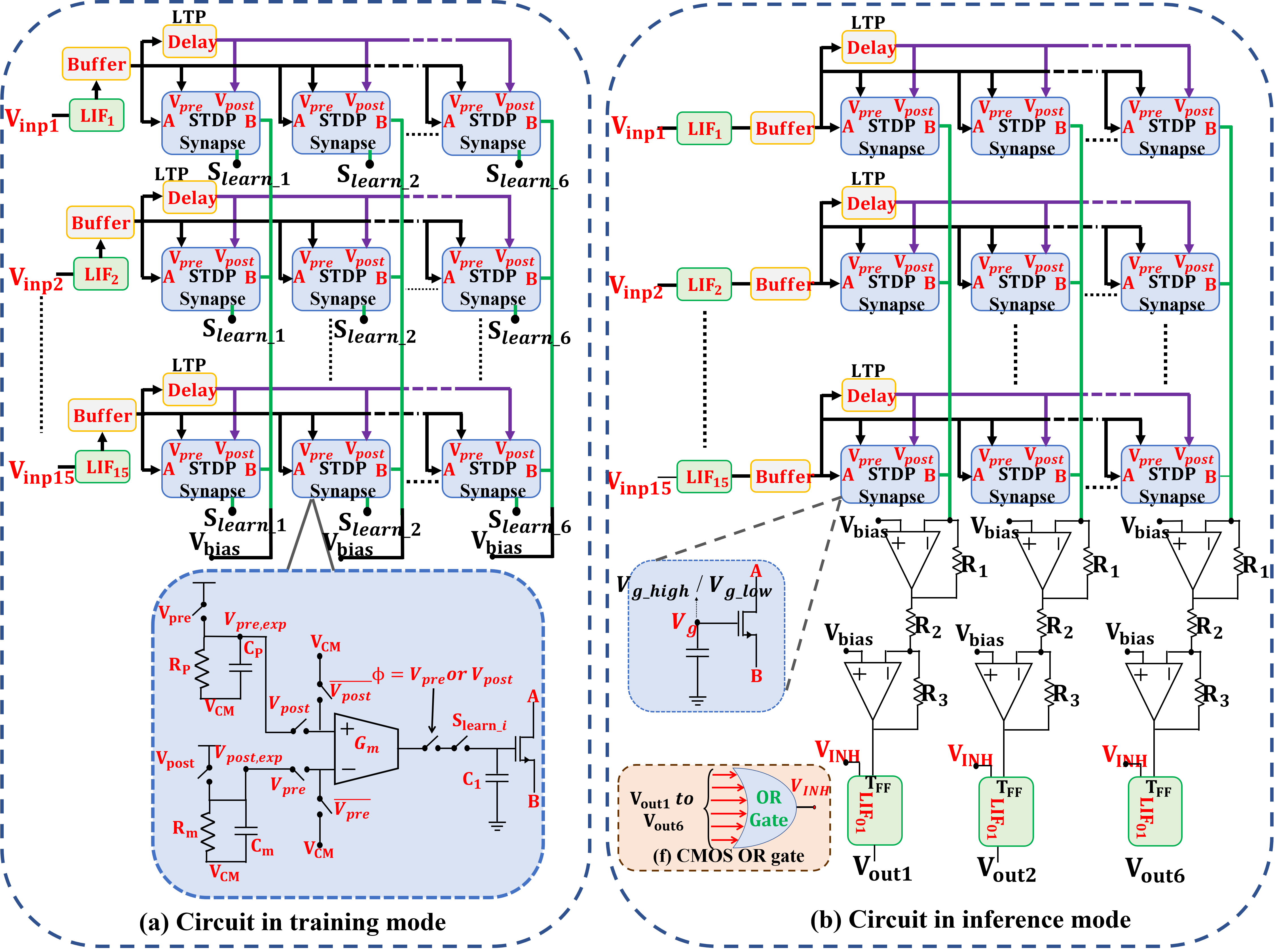"}
	\caption{The proposed system in (a) training mode and (b) inference mode.}
	\label{LTP_LTD}
\end{figure}

The long-term potentiation (LTP) obtained using the STDP learning circuit is used to update the synaptic weights according to the input. During the training phase, switch $S_{inference}$ is open and switch $S_{learn}$ is closed. 
Thus the output layer is disconnected from the synaptic crossbar resulting in the equivalent circuit shown in Fig.~\ref{LTP_LTD}(a). The B terminal of each STDP synapse is biased at a constant voltage. Each STDP circuit in the crossbar will receive presynaptic spikes from the input layer LIF neuron. The postsynaptic spikes to the STDP circuit will be a delayed version of the presynaptic spikes given from the input LIF neuron. Initially, all the synaptic weights are set as slow or at HRS. The process of updating weight based on the input can be divided into two different situations
\begin{enumerate}
	\item If the input is high (corresponding to a black pixel) or greater than the threshold voltage of LIF neurons, spikes will be generated by the input LIF neuron. These spikes will act as presynaptic spikes to the STDP circuit. The delay unit will send these spikes with the delay $\Delta t$ to the postsynaptic terminal of the STDP circuit. Each postsynaptic spike will precede presynaptic spike with the timing difference $\Delta t$. Thus, the weight change will monotonically increase with each postsynaptic spike leading to LTP.
	\item When the input is low (corresponding to a white pixel) or less than the threshold voltage of the LIF neuron, there will be no spike generated from the input LIF neuron, and the weight will remain unchanged or in HRS. 
\end{enumerate}
\begin{figure*}[!ht] 
	\centering
	\subfigure[]{\includegraphics[scale=0.36,trim={1.1cm 0 0 0},clip]{"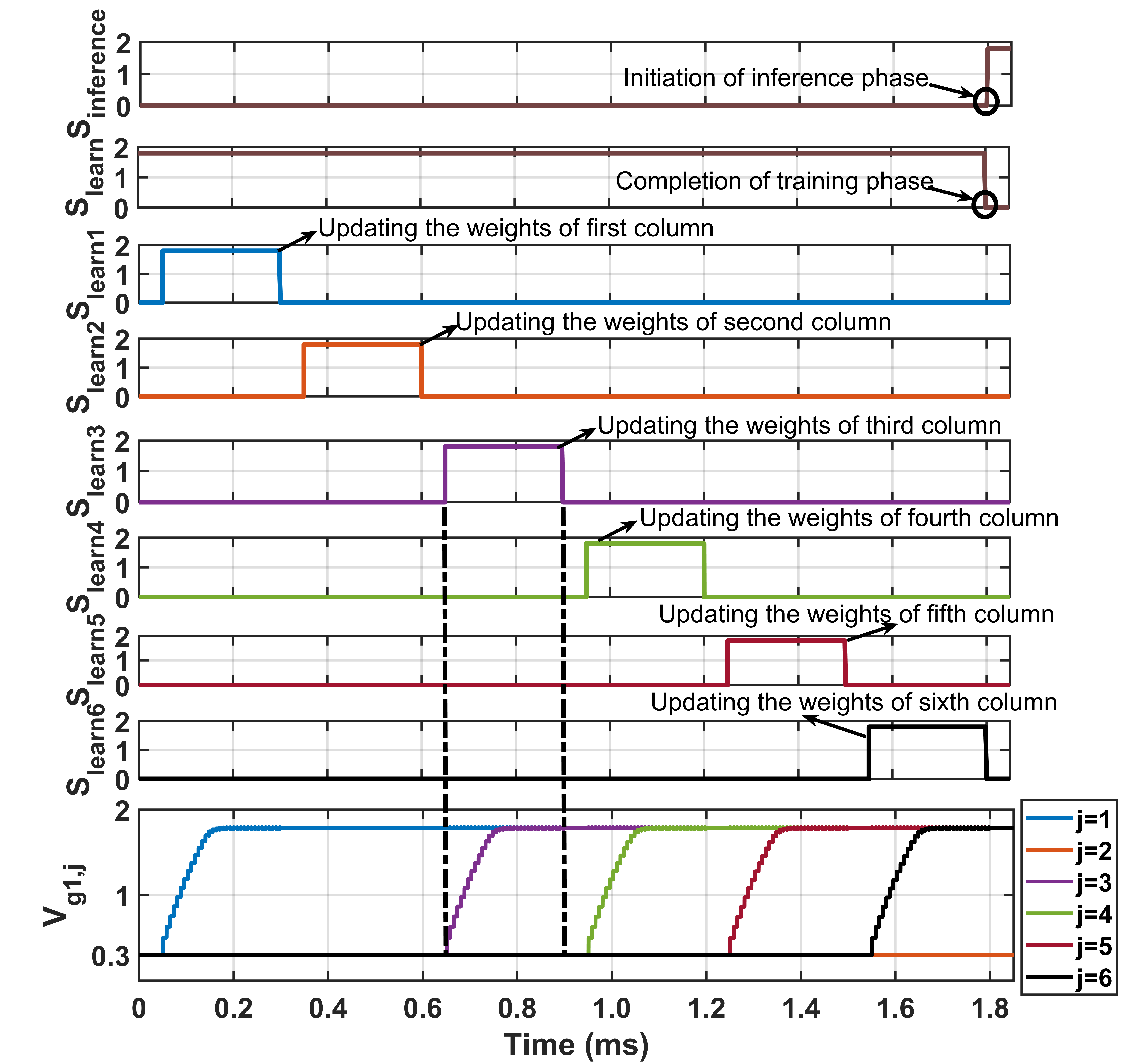"}}
	\subfigure[]{\includegraphics[scale=0.36]{"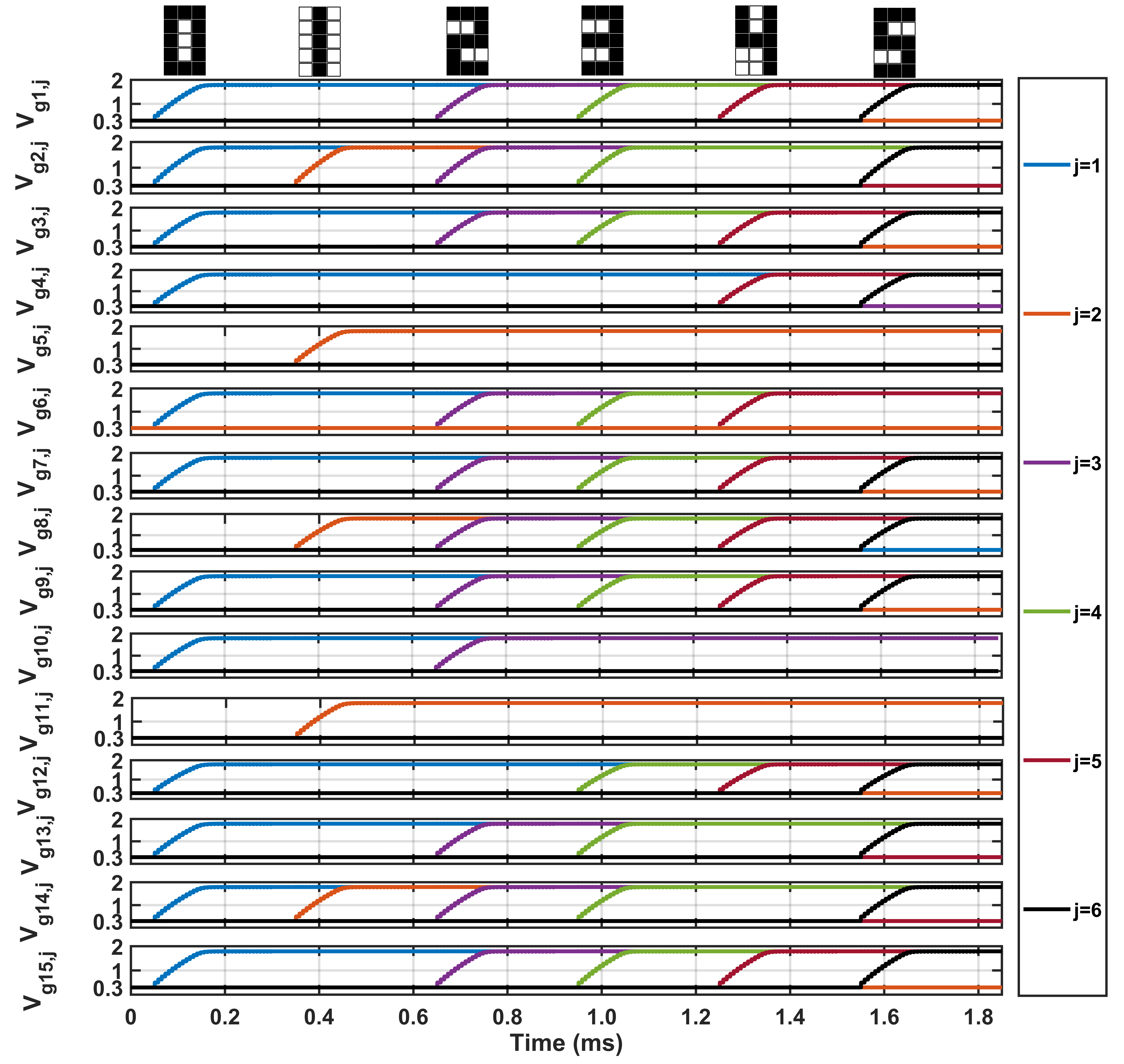"}}
	\caption{Simulation results of training phase (a) Control signals for training each column's synapses sequentially. (b) Synaptic state change of all STDP memristors corresponding to the input pattern. }
	\label{sim_train_infer}
\end{figure*}
\subsection{Circuit Design of Inference Phase}
During the inference phase, switch $S_{inference}$ is closed and switch $S_{learn}$ is open. The input layer is thus fully connected to the output layer through the synaptic crossbar, as shown in Fig~\ref{LTP_LTD}(b). The input corresponding to a pattern is given at the input neurons and processed with the synaptic crossbar. The weighted currents are summed column-wise and converted to a voltage using a summing amplifier ($A_{sum}$) and an inverter amplifier ($A_{inv}$). The output LIF neuron (i.e. the winner neuron) with the highest summed current will fire first and inhibit other neurons from firing. Unlike input layer LIF neurons, inhibitory interface (using $V_{INH}$) is required in output LIF neurons for implementing the winner-take-all (WTA) mechanism. Whenerever a winner neuron fires, the output of an OR gate turns high which resets other neurons through inhibitory terminals ($V_{INH}$).  
\section{Simulation Results}
This section presents the simulation results of the proposed CMOS based SNN system for the training and recognition phases. To validate the working of the proposed SNN circuit for pattern recognition, we have performed the image classification of six patterns (`0', `1', `2', `3', `4', `5') shown in Fig.~\ref{SNNsystem}(g). Each pattern is a binary image of 5 X 3 pixels, requiring 15 input LIF neurons, 6 output LIF neurons, and the crossbar of 90 STDP synapse circuits for classification. 
\subsection{Training Phase}
 The network is trained for each pattern sequentially and is in a configuration shown in Fig.~\ref{LTP_LTD}(a). Initially, all the STDP synapses are in the high resistance state (HRS). The control signals applied during the training phase is shown in Fig.~\ref{sim_train_infer}(a). All the synapses in the $j^{th}$ column will be trained in the $j^{th}$ learning cycle ($S_{learn}$\textunderscore $j$, $j$=1 $to$ 6). As an example shown in Fig.~\ref{sim_train_infer}(a), during high input of S$_{learn}$\textunderscore3 if input pattern `2' is given in the sequence (shown in Fig.~\ref{SNNsystem}(g)), the synaptic state of the STDP circuit in $1^{st}$ row, $3^{rd}$ column ($V_{g1,3}$) will increase according to STDP learning. Simultaneously, all other synaptic states of $3^{rd}$ column will be updated ($V_{gi,3}$, i=1 $to$ 15) as illustrated in Fig.~\ref{sim_train_infer}(b). 
  \begin{figure}[!t]
 	\centering
 	\includegraphics[scale=0.24,trim={2.0cm 0 2.0cm 0},clip]{"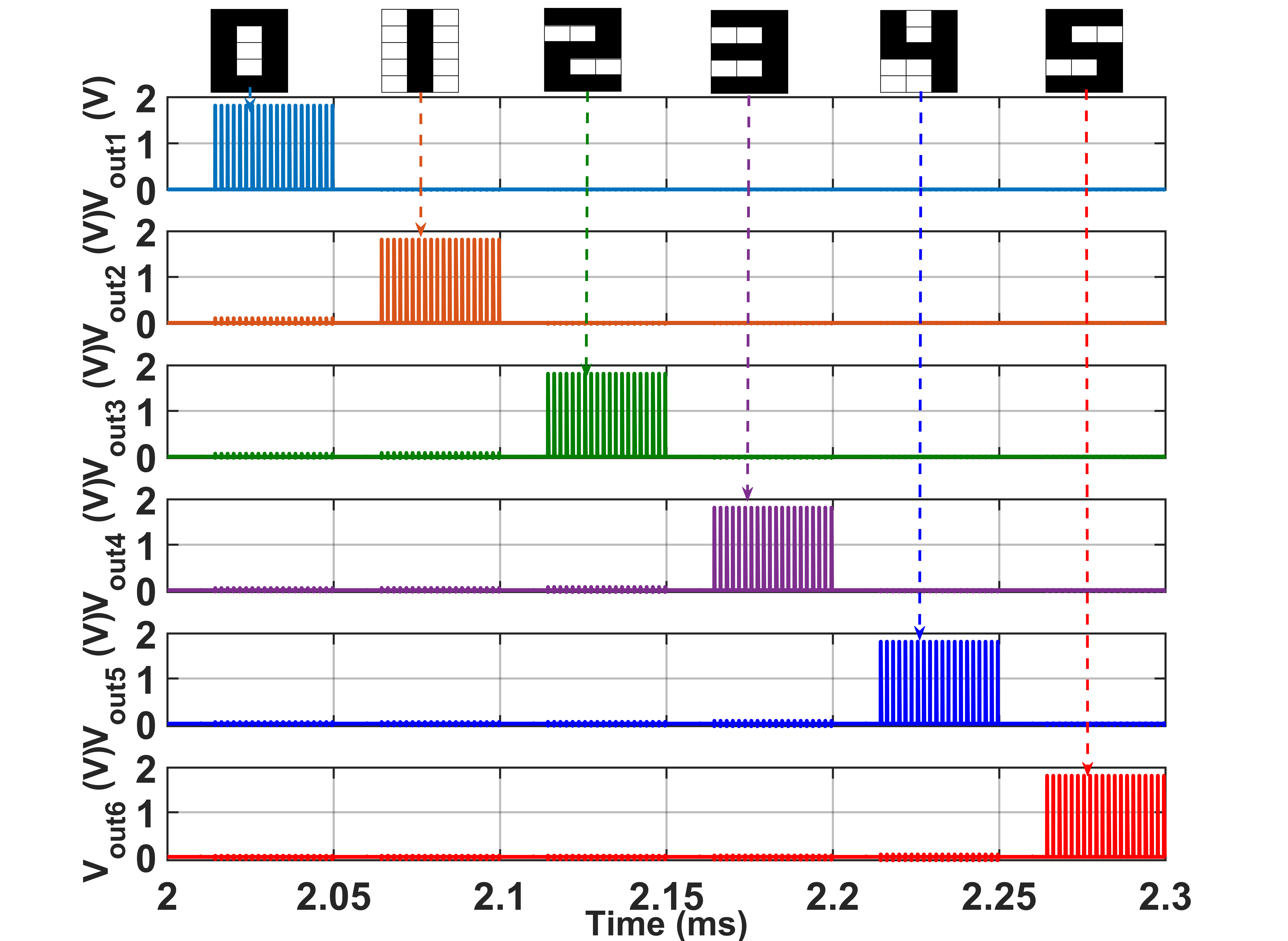"}
 	\caption{Simulation result of inference phase}
 	\label{inference}
 \end{figure} 
 The synaptic states ($V_{g1,3}$, $V_{g2,3}$, $V_{g3,3}$, $V_{g6,3}$, $V_{g7,3}$, $V_{g8,3}$, $V_{g9,3}$, $V_{g10,3}$, $V_{g13,3}$, $V_{g14,3}$, $V_{g15,3}$) emulated by the gate voltage of the floating transistor of the STDP circuit (refer to Fig.~\ref{STDP}) increases (or the synaptic resistance decreases to LRS) while remaining states are in their initial HRS. The change in the synaptic states of other columns in j$^{th}$ learning cycle corresponding to the given pattern is also shown in Fig.~\ref{sim_train_infer}(b). The process of changing the synaptic weights of each column sequentially according to the given pattern shows the unsupervised in-situ STDP learning of the network.
\subsection{Inference Phase}
After training the network for all patterns, each synaptic crossbar column ($j$=1 to 6) will store the weights according to the corresponding pattern (0 to 5). Each pattern is presented to the network for 10 $\mu$s. Only the winner neuron will fire when its corresponding pattern is given at the input while inhibiting the other neurons. Fig.~\ref{inference} shows the firing response of the output LIF neurons, which confirms that each output neuron is able to recognise the corresponding pattern.
\subsection{Robustness}
\begin{figure}[!t]
	\centering
	\includegraphics[scale=0.24,trim={0 0 0 0},clip]{"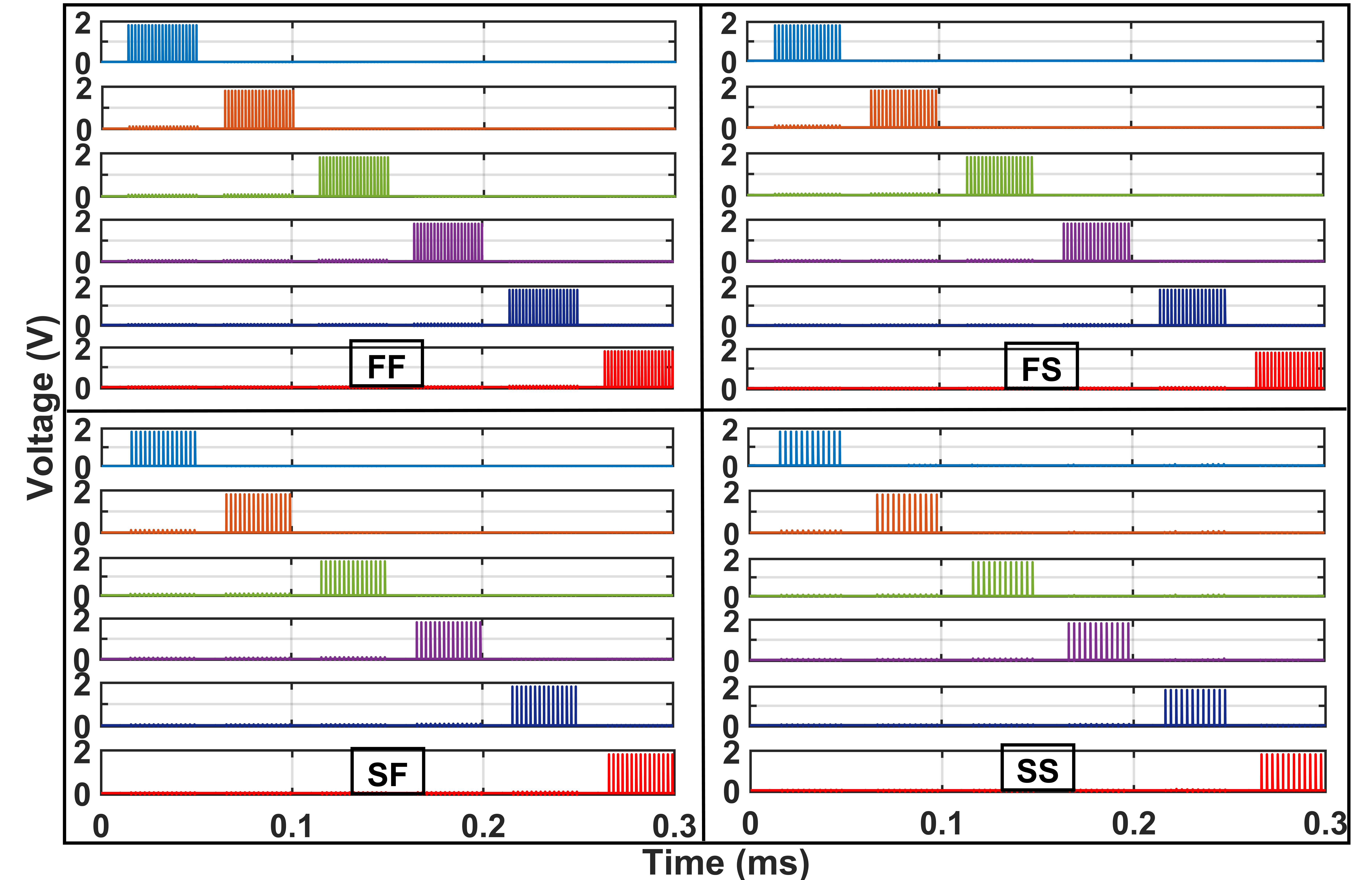"}
	\caption{Response of output neurons in recognition phase for all extreme process corners at temperature = $27^{\circ}$C.}
	\label{PVT}
\end{figure}

  In order to analyse the robustness of the proposed SNN system for pattern recognition of digits (`0', `1', `2', `3', `4', `5'), we have evaluated the recognition rate of the circuit for different noisy patterns. Also, we have considered the limitations of CMOS fabrication technology by adding process and temperature variations of all the circuit components in the architecture. The noise is added by inverting the pixel of the original pattern from 0 to 1 and vice versa in the percentage of 6.67\%, 13.3\% and 20\% as done in \cite{N1}. We have tested all possible combinations of noisy patterns listed in Table II. Taking the case of 13.3\% noise patterns, i.e. change of 2 pixels in 5$\times$3 pixel image, there are 105 cases of noisy patterns for one digit patterns and 630 cases for six digit patterns. Response of output neuron in inference phase shown in Fig.~\ref{PVT} verifies the correctness of the circuit under extreme process corners. Each case of noisy pattern is also tested for process corners (SS, SF, FS, FF, TT) and different temperatures ($0 ^{\circ}C, 27 ^{\circ}C, 50 ^{\circ}C, 100 ^{\circ}C$). As shown in Table II, the recognition rate of the circuit-based simulations matches closely with the analytical results using MATLAB. This proves the robustness of the circuit for noisy patterns in the presence of process and temperature variation. 
 \begin{table}[t]
 	\renewcommand{\thefootnote}{\fnsymbol{footnote}}
 	\caption{Recognition rate for various noisy patterns.}
 	\label{table}
 	\centering
 	\begin{tabular}{|p{0.75cm}|p{1.75cm}|p{2.2cm}|p{2cm}|}
 		\hline
 		\textbf{Noise (\%)} & \textbf{Total noisy patterns} & \textbf{Expected recognition rate from analytical results(\%)} & \textbf{Recognition rate from circuit simulation(\%)}\\
 		\hline
 		\textbf{0} & 1 $\times$ 6 = 6 & 100 & 100\\
 		\hline
 		\textbf{6.67} & 15 $\times$ 6 = 90 & 77.8 & 77.8\\
 		\hline
 		\textbf{13.3} & 105 $\times$ 6 = 630 & 55.4 & 55.4\\
 		\hline
 		\textbf{20.0} & 455$\times$ 6 = 2730 & 46.2 & 46.2\\
 		\hline
 	\end{tabular}	
 \end{table}

 \begin{table*}[!ht] \label{t1}
 	\renewcommand{\thefootnote}{\fnsymbol{footnote}}
 	\caption{Comparison of Circuit Implementation of Different Memristive Crossbar-Based Neural Network Circuits}
 	\label{table}
 	\centering
 	\begin{tabular}{|p{1.83cm}|p{1.78cm}|p{1.7cm}|p{3.2cm}|p{1.7cm}|p{2.9cm}|p{1.91cm}|}
 		\hline
 	\cellcolor{Gray}\multirow{3}{*}{} & \cellcolor{Gray}\bfseries Shamshi et al., & \cellcolor{Gray}\bfseries X. Wu et al., &\cellcolor{Gray} \bfseries V. Saxena et al., &\cellcolor{Gray} \bfseries Z. Chen et al., &\cellcolor{Gray} \bfseries M. Chu et al., & \cellcolor{Gray}\\
 	\cellcolor{Gray}	{} &\cellcolor{Gray} \bfseries TVLSI, &\cellcolor{Gray} \bfseries JETCAS, &\cellcolor{Gray} \bfseries ISCAS, &\cellcolor{Gray} \bfseries TVLSI, &\cellcolor{Gray} \bfseries TIE, &\cellcolor{Gray} \bfseries This work \\
 	\cellcolor{Gray}	{} &\cellcolor{Gray} \bfseries 2018 \cite{a3} &\cellcolor{Gray} \bfseries 2015 \cite{a4} &\cellcolor{Gray}\bfseries 2018 \cite{a6} &\cellcolor{Gray} \bfseries 2021 \cite{a1} &\cellcolor{Gray} \bfseries 2015 \cite{46} &\cellcolor{Gray}  \\
  \hline
 	
 		\cellcolor{Gray} \textbf{Type of memristor} & Yakopcic SPICE model & Yakopcic SPICE model & STDP synapse circuit & Biolek SPICE model & PCMO based fabricated memristive device & CMOS STDP synapse circuit \\
 		\hline
 		\cellcolor{Gray} \textbf{Learning rule} & STDP, ex-situ & STDP, in-situ & STDP, in-situ &  Hebbian, in-situ & Modified STDP learning &  STDP, in-situ\\
 		\hline
 		\cellcolor{Gray}\textbf{Type of data for updating memristor weight} & Programming memristor using digital pulses &Spikes & Spikes & Digital pulses of \textpm 5 V & Digital pulses of \textpm 3 V  & Spikes \\
 		\hline
 		\cellcolor{Gray}\textbf{Learning type} & Supervised & Supervised & Supervised & Unsupervised & Unsupervised & Unsupervised \\
 		\hline
 		\cellcolor{Gray}\textbf{WTA mechanism} & - & Bus interface circuit made of digital gates and D-FF & Bus interface circuit made of digital gates and D-FF & Using PMOS transistors with output layer neurons & Control logic & Using a CMOS OR gate \\
 		\hline
 		\cellcolor{Gray}\textbf{Image size/no. of patterns} & (5 $\times$ 5) / 4 & (8 $\times$ 8) / 10 & (8 $\times$ 8) / 10 & (5 $\times$ 3) / 3 & (5 $\times$ 6) / 10 & (5 $\times$ 3) / 6\\
 		\hline
 		\cellcolor{Gray}\textbf{No. of synapse in crossbar} & 100 & 640 & 640 & 90 (2 memristors per synapse) &300 & 90\\
 		\hline
 		\cellcolor{Gray}\textbf{Full CMOS transistor level circuit implementation} & Not implemented (used memristor model) & Not implemented (used memristor model)  & Not implemented (simulated using macro-models memristive synapse and WTA neuron macro-models using Brian2 libraries in python & Not implemented & Hardware implementation using memristor crossbar array in one chip, FPGA and LIF neurons integrated in another chip & Full CMOS transistor level circuit implementation of system \\
 		\hline
 	\end{tabular}	
 \end{table*}  
\subsection{Comparison and Discussion}
This paper presents the full CMOS-based transistor level implementation of SNN for pattern recognition which makes it different from other existing literature listed in Table III. The proposed SNN system uses more bio-plausible mechanisms, including unsupervised STDP in-situ learning of CMOS memristive synapses. The employment of LIF neurons \cite{a3} in the output layer, which have an inhibitory interface, has aided in achieving the WTA mecha	nism discussed in section IV-B in a more straightforward manner. Authors in \cite{a3} have proposed the hardware architecture using spiking LIF neurons and synaptic circuits. However, they have used the SPICE model of the memristor in their simulations. Also, the weights were calculated in an ex-situ fashion and were mapped to the memristive crossbar using digital pulses. This makes the use of an extra circuit to read/write a weight value from/to a synaptic circuit. In view of this, our work implements in-situ learning avoiding the use of any additional circuit overhead. In paper \cite{a4,a6}, SNN is designed for digit recognition using in-situ supervised STDP learning. Supervised learning requires a teacher signal for each output neuron. However, in our designed SNN circuit for digit recognition, no interference of output neuron or teacher signal is required. The WTA mechanism implemented in \cite{a4,a6} requires a bus interface circuit made of digital gates and D-FFs. In contrast, our system uses only a CMOS OR gate with output neurons to perform the inhibition. In \cite{a6}, CMOS based memristor circuit with STDP learning is proposed but simulated using memristive synapse and WTA neuron macro models using brian2 libraries in Python. Work \cite{a1} implements the non-spiking memristive neural network for pattern recognition using Hebbian in-situ learning. They have used the SPICE model of memristor and programmed it using high magnitude bipolar digital pulses. Also, their architecture requires two memristors per synapse. In contrast, our work uses a more bio-plausible way of implementing the pattern recognition application using SNN with STDP learning and LIF neuron at input and output. In \cite{46}, instead of taking any SPICE model of the memristor, a fabricated PCMO based memristive device is used. However, full SNN hardware was not implemented in one chip. Also, it requires the use of FPGA as a controlling unit. In comparison, our proposed work does not require FPGA and gives the on-chip design of SNN for pattern recognition.

\section{Prospect}

Although the proposed CMOS SNN system is used for pattern recognition of six digits, it can also be used for visual pattern recognition. The complete neuromorphic system for visual pattern recognition will contain CMOS photoreceptor which converts optical data into a spike train. The generated spike train will be input to the memristive neural network for recognition.

In view of the CMOS STDP circuit used in this paper, we have also applied it to demonstrate the heart rate classification. We have explored this circuit to obtain Beinenstock-Cooper-Munro (BCM) characteristics for rate-based learning other than timing based learning. Unlike spike-timing (STDP) based learning, BCM learning modifies the synaptic weights based on pre and postsynaptic spike frequencies. It has been reported that by limiting the interaction of pre and postsynaptic spikes to the nearest-neighbour spike interaction only, BCM learning can be replicated from pair-based STDP learning circuit \cite{bcm1,bcm2}. The synaptic weight change is determined by the threshold frequency $\theta$, if postsynaptic firing rate $f_x< \theta$, depression occurs (change in synaptic weight is negative), if firing rate $f_x> \theta$, potentiation occurs (change in synaptic weight is positive). The threshold frequency between potentiation and depression (at which the rate change of the synaptic weight is zero) is given by \eqref{ref9}, where $A_+,  A_-, \tau_+, \tau_- $ are the parameters of the STDP curve \cite{bcm1};   
\begin{equation} \label{ref9}
\theta=-\frac{A_+/\tau_- + A_-/\tau_+ }{A_++A_-}
\end{equation}
\begin{table}[t]
	\renewcommand{\thefootnote}{\fnsymbol{footnote}}
	\caption{Parameters Used for Different Threshold BCM Learning}
	\label{table}
	\centering
	\begin{tabular}{|p{1.2cm}|p{2.2cm}|p{2.2cm}|}
		\hline
		\textbf{Parameter} & \textbf{$\theta_1$ (For 60 BPM)} & \textbf{$\theta_2$ (For 120 BPM)}\\
		\hline
		\textbf{$A_+$} & 0.267 & 0.19\\
		\hline
		\textbf{$\tau_+$} & 0.7 & 0.7 \\
		\hline
		\textbf{$A_-$} & 0.175 & 0.138\\
		\hline
		\textbf{$\tau_-$} & 1.7 & 1.7\\
		\hline
	\end{tabular}	
\end{table}

\begin{figure}[!t]
	\centering
	\includegraphics[scale=0.32]{"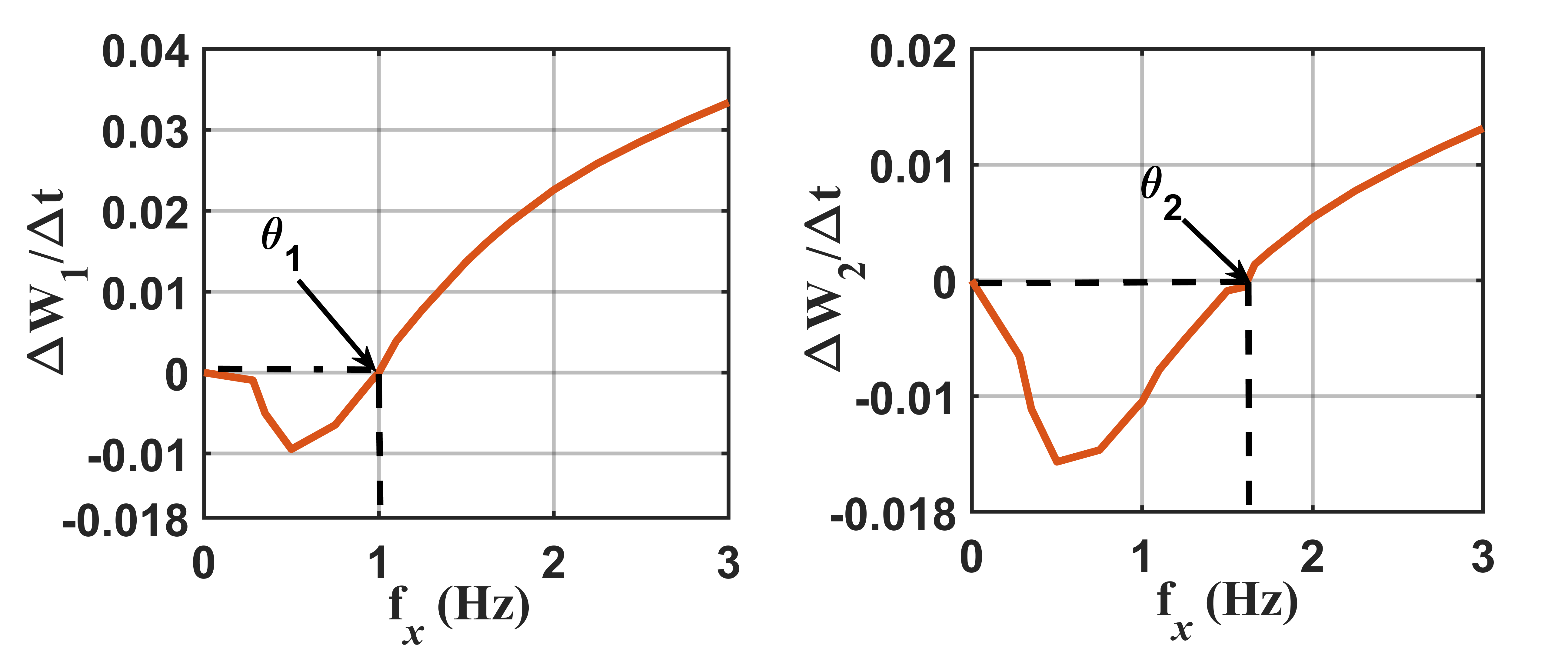"}
	\caption{BCM learning curve of the nearest-neighbour pair-based STDP circuit with different thresholds ($\theta_1=1 Hz, \theta_2=1.667 Hz$)}
	\label{BCMcurve}
\end{figure}
\begin{table}[!t]
	\renewcommand{\thefootnote}{\fnsymbol{footnote}}
	\caption{Heart Rate Clasification output for ECG Database}
	\label{table}
	\centering
	\begin{tabular}{|p{0.3cm}|p{1.6cm}|p{1.7cm}|p{1.6cm}|p{1.3cm}|}
		\hline
		\textbf{S. No.} & \textbf{Actual Heart Rate (bpm)} & \textbf{$\Delta w_1$(For 60 bpm)} & \textbf{$\Delta w_2$(For 100 bpm)} & \textbf{Classifier Output}\\
		\hline
		\textbf{1} & 70.36 & 5.783$*10^{-3}$ & -6.416$*10^{-3}$ & NORMAL \\
		\hline
		\textbf{2} & 82.15 & 1.073$*10^{-2}$ & 	-2.966 $*10^{-3}$ & NORMAL \\
		\hline
		\textbf{3} & 87.42 & 1.275$*10^{-2}$ & -1.546$*10^{-3}$ & NORMAL \\
		\hline
		\textbf{4} & 96.84 & 1.601$*10^{-2}$ & 	-7.544 $*10^{-4}$ & NORMAL \\
		\hline
		\textbf{5} & 56.00 & -1.642$*10^{-3}$ & -1.154$*10^{-2}$ & LOW \\
		\hline
		\textbf{6} & 43.06 & -7.143$*10^{-3}$ & -1.501$*10^{-2}$ & LOW \\
		\hline
		\textbf{7} & 134.87 & 2.582$*10^{-2}$ & 7.722$*10^{-3}$ & HIGH \\
		\hline
		\textbf{9} & 108.42 & 1.952$*10^{-2}$ &  3.239 $*10^{-3}$ & HIGH \\
		\hline
		\textbf{1} & 106.72 & 1.903$*10^{-2}$ & 2.895$*10^{-3}$ & HIGH \\
		\hline
		\textbf{10} & 51.2 & -3.708$*10^{-3}$ & -1.288$*10^{-2}$ & LOW \\
		\hline
	\end{tabular}	
\end{table}
In rested condition, the lower and the upper safe limits of heart rate (in beats per minute) are 60 bpm  and 100 bpm, which corresponds to 1 Hz and 1.667 Hz respectively. The two BCM curves corresponding to $\theta_1,\theta_2$ are shown in Fig.~\ref{BCMcurve}, which are obtained by optimising the two STDP circuits for parameters listed in Table IV. The classifier (containing two BCM learning enabled STDP circuits) has been tested with ECG datasets taken from PhysioBank ATM. $\Delta W_1$, $\Delta W_2$ correspond to the weight change observed in STDP circuit tuned for $\theta_1 = 1$ Hz and $\theta_2 = 1.667 $ Hz respectively. The input spike data is generated in MATLAB by detecting the Q peaks from the ECG signal. Based on the rate of weight change the proposed heart rate classifier correctly classifies the ECG database (present on PhysioBank ATM) as shown in Table V. The classified rate is normal if potentiation is seen in  $\Delta W_1$ (positive value of $\Delta W_1$) and depression in $\Delta W_2$ (negative value of $\Delta W_2$). 
\section{Conclusion}
This paper presents the complete CMOS based transistor-level implementation of SNN for pattern recognition. It integrates 15 input CMOS LIF neurons, a crossbar array of 90 CMOS memristive synapse circuits and 6 output LIF neurons for classification of six 5$\times$3 pixel images. This system embeds on-chip learning and inference using STDP mechanism and WTA property. The designed CMOS based SNN system is validated with the training and inference simulation results for the classification of six patterns. The proposed circuit is shown to be robust for process and temperature variations. Moreover, the recognition rate obtained from the circuit simulation under noisy patterns matches the analytical results, proving the accuracy of the proposed CMOS-based SNN circuit. Other than the timing based STDP learning, rate-based learning is also explored using the same pair-based STDP circuit and shown its application for heart rate classification by obtaining the BCM characteristics of the STDP circuit.


%
\section*{Acknowledgment}
The authors thank Ministry of Education (MoE), MeitY and SCL, Govt. of India for providing institute fellowship, availing the tools through SMDP-C2SD project and 180 nm PDK, respectively.

\ifCLASSOPTIONcaptionsoff
  \newpage
\fi

\end{document}